\documentclass{PoS}

\title{Gauge theories and gravity}

\ShortTitle{Gauge theories and gravity}

\author{\speaker{R.~F.~Sobreiro}\\
        UFF $-$ Universidade Federal Fluminense, Instituto de F\'{\i}sica, Campus da Praia Vermelha, Avenida General Milton Tavares de Souza s/n, 24210-346, Niter\'oi, RJ, Brasil.\\
        E-mail: \email{sobreiro@if.uff.br}}

\author{A.~A.~Tomaz\\
        UFF $-$ Universidade Federal Fluminense, Instituto de F\'{\i}sica, Campus da Praia Vermelha, Avenida General Milton Tavares de Souza s/n, 24210-346, Niter\'oi, RJ, Brasil.\\
        E-mail: \email{tomaz@if.uff.br}}

\author{V.~J.~Vasquez Otoya\\
        IFSEMG $-$ Instituto Federal de Educa\c{c}\~ao, Ci\^encia e Tecnologia, Rua Bernardo Mascarenhas 1283, 36080-001, Juiz de Fora, MG, Brasil.\\
        E-mail: \email{victor.vasquez@ifsudestemg.edu.br}}

\abstract{Pure gauge theories for de Sitter, anti de Sitter and orthogonal groups, in four-dimensional Euclidean spacetime, are studied. It is shown that, if the theory is asymptotically free and a dynamical mass is generated, then an effective geometry may be induced and a gravity theory emerges.}

\FullConference{7th Conference Mathematical Methods in Physics - Londrina 2012,\\
		16 to 20 April 2012\\
		Rio de Janeiro, Brazil}

\begin{document}

\section{Introduction}\label{intro}

Although general relativity and the standard model are different in nature, is widely known that gravity can be viewed as a gauge theory \cite{Utiyama:1956sy,Kibble:1961ba,Zanelli:2005sa}. To do that, two fundamental fields are introduced, the vierbein $e$ and the spin connection $\omega$. The geometric properties of spacetime are obtained from specific composite fields that are constructed using these fundamental fields \cite{Zanelli:2005sa}. It turns out that the deep relation between the fields of gravity and spacetime ruins the possibility of a quantum description of gravity independent of the background geometry, \emph{i.e.}, a quantum field should not depend on parameters that also fluctuate. Moreover, even in a background dependent quantization, the Einstein-Hilbert action itself is not enough to ensure perturbative quantum stability of gravity \cite{'tHooft:1974bx}. To circumvent these problems, by generalizing the gauge groups and their respective actions, many other theories have been proposed. For instance, as discussed in \cite{MacDowell:1977jt,Stelle:1979aj,Mahato:2004zi,Tresguerres:2008jf,Mielke:2010zz}, spontaneous symmetry breaking based on a Higgs-like mechanism was used to make the vierbein emerge. In these works, besides de Sitter groups, several groups were considered, as well as different starting actions that encode gravity as a limit.

The present work is about de Sitter gauge theories in four-dimensional Euclidean spacetime. The starting action is the massless pure Yang-Mills action with $SO(m,n)$ gauge symmetry where $m+n=5$ and $m\in\{0,1,2\}$. Thus, renormalizability is ensured at least to all orders in perturbation theory. The choice of a Euclidean space is not accidental; it follows from the fact that any quantum field theory is actually treatable only in Euclidean spaces (even perturbatively, where a Wick rotation is needed for reliable quantum computations). Moreover, in a Euclidean manifold, space and time are indistinguishable, and thus, time evolution of any physical system becomes, at least, unclear. On the other hand, non-Abelian gauge theories have two main effects. First, the theory is perturbatively asymptotically free \cite{Gross:1973id,Politzer:1973fx}. Second, dynamical mass parameters might arise at non-perturbative level as the coupling parameter increases \cite{Dudal:2005na,Dudal:2011gd}. The combination of both effects can be used to show that an induced gravity theory can emerge naturally, where the running parameters induce a dynamical symmetry breaking to Lorentz type groups. Then, a suitable mapping enables a gravity theory to rise. In this theory, the dynamical mass plays the fundamental role of distinguishing the quantum and classical sectors of gravity. The quantum sector is a standard spin-1 gauge theory, and the classical sector is an effective geometrodynamics \cite{Sobreiro:2011hb}.

\section{de Sitter gauge theories and effective geometry}

The gauge group $SO(m,n)$ defines an internal flat space $\mathbb{R}^{m,n}_S$ which has no relation to the four dimensional spacetime $\mathbb{R}^4$. The $10$ anti-hermitian generators $J^{AB}$ of the gauge group, where $\{A,B,\ldots\}\in\{5,0,1,2,3\}$, are antisymmetric in their indices. The invariant Killing metric is $\eta^{AB}\equiv\mathrm{diag}(\epsilon,\varepsilon,1,1,1)$ where $\epsilon=(-1)^{(2-m)!}$ and $\varepsilon=(-1)^{m!+1}$. The group can be decomposed by projecting the fifth coordinate, $SO(m,n)\equiv SO(m!-1,n)\otimes S(4)$, in such a way that
\begin{eqnarray}
\left[J^{ab},J^{cd}\right]&=&-\frac{1}{2}\left[\left(\eta^{ac}J^{bd}+\eta^{bd}J^{ac}\right)-\left(\eta^{ad}J^{bc}+\eta^{bc}J^{ad}\right)\right]\;,\nonumber\\
\left[J^a,J^b\right]&=&-\frac{\epsilon}{2}J^{ab}\;,\nonumber\\
\left[J^{ab},J^c\right]&=&\frac{1}{2}\left(\eta^{ac}J^b-\eta^{bc}J^a\right)\;,\label{alg2}
\end{eqnarray}
where, $\{a,b,\ldots\}\in\{0,1,2,3\}$ and $J^{5a}=J^a$ and $\eta^{ab}\equiv\mathrm{diag}(\varepsilon,1,1,1)$.

The fundamental field is the 1-form gauge connection, an algebra-valued quantity in the adjoint representation, $Y=Y^A_{\phantom{A}B}J_A^{\phantom{A}B}=A^a_{\phantom{a}b}J_a^{\phantom{a}b}+\theta^aJ_a$, whose gauge transformation is given by $Y\longmapsto u^{-1}\left(\kappa^{-1}\mathrm{d}+Y\right)u\;\Big|\;u\in SO(m,n)$, where, obviously, $\kappa$ is a dimensionless coupling parameter and $\mathrm{d}$ the exterior derivative. At infinitesimal level, we have $Y\longmapsto Y+\nabla\zeta$,  where $u=\exp{(\kappa\zeta)}\approx1+\kappa\zeta$ and $\nabla=\mathrm{d}+\kappa Y$ is the full covariant derivative. This transformation decomposes as
\begin{eqnarray}
A^a_{\phantom{a}b}&\longmapsto& A^a_{\phantom{a}b}+\mathrm{D}\alpha^a_{\phantom{a}b}-\frac{\epsilon\kappa}{4}\left(\theta^a\xi_b-\theta_b\xi^a\right)\;,\nonumber\\
\theta^a&\longmapsto&\theta^a+\mathrm{D}\xi^a+\kappa\alpha^a_{\phantom{a}b}\theta^b\;,\label{gt2}
\end{eqnarray}
where $\zeta=\alpha^a_{\phantom{a}b}J_a^{\phantom{a}b}+\xi^aJ_a$ and $\mathrm{D}=\mathrm{d}+\kappa A$ is the covariant derivative with respect to the sector $SO(m!-1,n)$. The 2-form field strength is obtained from $F=\nabla^2=\mathrm{d}Y+\kappa YY$, which decomposes as $F=\left(\Omega^a_{\phantom{a}b}-\frac{\epsilon\kappa}{4}\theta^a\theta_b\right)J_a^{\phantom{a}b}+K^aJ_a$ where $\Omega^a_{\phantom{a}b}=\mathrm{d}A^a_{\phantom{a}b}+\kappa A^a_{\phantom{a}c}A^c_{\phantom{c}b}$ and $K^a=\mathrm{D}\theta^a=\mathrm{d}\theta^a-\kappa A^a_{\phantom{b}b}\theta^b$.

It turns out that, the most general, gauge invariant, massless, renormalizable action is the usual Yang-Mills action. This action can be written as
\begin{equation}
S_{\mathrm{YM}}=\frac{1}{2}\int\left[\Omega^a_{\phantom{a}b}{*}\Omega_a^{\phantom{a}b}+\frac{1}{2}K^a{*} K_a-\frac{\epsilon\kappa}{2}\Omega^a_{\phantom{a}b}{*}(\theta_a\theta^b)+\frac{\kappa^2}{16}\theta^a\theta_b{*}(\theta_a\theta^b)\right]\;,\label{ym0}
\end{equation}
where ${*}$ denotes the Hodge dual operation in spacetime. Besides quantum stability, this action has three main properties. First, the theory is asymptotically free \cite{Gross:1973id,Politzer:1973fx}. As a consequence, a non-pertubative behavior is expected at the infrared regime, which becomes more evident by means of an increasing of the coupling parameter $\kappa$. Second, the non-linearity of the theory also favors the condensation of composite operators and thus the possibility of dynamical mass parameters to emerge \cite{Dudal:2005na,Dudal:2011gd}. Third, on the other hand, at least one mass parameter is required for quantization improvements in order to fix the so called Gribov ambiguities \cite{Dudal:2005na,Dudal:2011gd}.

At the ultra-violet regime, an important feature of the present action is the absence of mass parameters. Usually, in de Sitter gravity \cite{Stelle:1979aj,Tresguerres:2008jf}, the field $\theta^a$ possesses components $\theta^a_\mu$ that carry ultra-violet (UV) dimension $0$ and always appear with a mass scale factor (the cosmological constant) to adjust the correct UV dimension of a connection component. In the present model, the components $\theta^a_\mu$ carry UV dimension 1 and then cannot be directly associated with coframes. The realization of such identification is only possible at the IR regime where dynamical masses emerge. The first step in this achievement is to assume, independently of the physical mechanism, the existence of a mass scale, denoted here by $\gamma$. The existence of a mass allows a rescaling of the fields $A \longmapsto \kappa^{-1}A$ and $\theta \longmapsto \kappa^{-1}\gamma\theta$. In this rescaling, the mass parameter affects only the $\theta$-sector, transforming it in a field with dimensionless components. The consequence for the action (\ref{ym0}) is
\begin{equation}
S=\frac{1}{2\kappa^2}\int\left[\overline{\Omega}^a_{\phantom{a}b}{*}\overline{\Omega}_a^{\phantom{a}b}+\frac{\gamma^2}{2}\overline{K}^a{*}\overline{K}_a-\frac{\epsilon\gamma^2}{2}\overline{\Omega}^a_{\phantom{a}b}{*}(\theta_a\theta^b)+\frac{\gamma^4}{16}\theta^a\theta_b{*}(\theta_a\theta^b)\right]\;,\label{ym1}
\end{equation}
where $\overline{\Omega}^a_{\phantom{a}b}=\mathrm{d} {A}^a_{\phantom{a}b}+ {A}^a_{\phantom{a}c} {A}^c_{\phantom{c}b}$, $\overline{K}^a=\mathrm{D}\theta^a$ and the covariant derivative is now $\mathrm{D}=\mathrm{d}+A$. 

The consequence for the Lie algebra is that the second of Eq.~(\ref{alg2}) is replaced by $\left[J^a,J^b\right]=-\frac{\epsilon\gamma^2}{2\kappa^2}J^{ab}$. Thus, assuming that, at low energies, the quantity $\gamma^2/\kappa^2$ is very small for some non-perturbative scales, an In\"on\"u-Wigner contraction takes place \cite{Inonu:1953sp}. The result is that the de Sitter group is contracted down to the Poincar\'e group where $J^a\longmapsto P^a$. However, this contraction induces a symmetry breaking of the action in Eq.~(\ref{ym1}), not to the Poincar\'e group $ISO(m!-1,n)$, but to the Lorentz group $SO(m!-1,n)$. This is evident if one realizes that the Poincar\'e group is not a subgroup of the de Sitter group, and the Lorentz group is a stability subgroup with respect to the de Sitter group. Under this dynamical breaking, the gauge transformations in Eq.~(\ref{gt2}) reduce to $A^a_{\phantom{a}b} \longmapsto {A}^a_{\phantom{a}b}+\mathrm{D}\alpha^a_{\phantom{a}b}$ and $\theta^a \longmapsto \theta^a-\alpha^a_{\phantom{a}b}\theta^b$. Thus, the field $\theta$ migrates to the matter sector while $A$ is a gauge connection for the Lorentz group.

To associate the action in Eq.~(\ref{ym1}) with gravity \cite{Sobreiro:2011hb,Sobreiro:2010ji,Sobreiro:2012book}, it is necessary to define an isomorphism that maps each point $x\in\mathbb{R}^4$ into a point $X\in\mathbb{M}^4$, the latter being the effective deformed spacetime. The local gauge group $SO(m!-1,n)$ defines, at each point $X$, the isometries of the tangent space $T_X(\mathbb{M})$. It is also convenient to impose that the space of $p$-forms in $\mathbb{R}^4$ is identified with the space of $p$-forms in $\mathbb{M}^4$, and the same for the Hodge duals, \emph{i.e.}, $\Pi^p\longmapsto\widetilde{\Pi}^p$ and $\ast\Pi^p\longmapsto\star\widetilde{\Pi}^p$, where $\star$ is the Hodge dual in $\mathbb{M}^4$. Moreover, $\theta$ and $A$ can be identified with the vierbein $e$ and spin connection $\omega$, \emph{i.e.}, $\omega^{\mathfrak{ab}}_\mu(X)dX^\mu=\delta^{\mathfrak{a}}_a\delta^{\mathfrak{b}}_bA^{ab}_\mu(x)dx^\mu$ and $e^{\mathfrak{a}}_\mu(X)dX^\mu=\delta_a^{\frak{a}}\theta^a_\mu(x)dx^\mu$. The indices $\{\mathfrak{a},\mathfrak{b},\ldots\}$ belong to the tangent space $T_X(\mathbb{M})$. Thus, each gauge configuration $(A,\theta)$ is identified with an effective geometry $(\omega,e)$, and gravity emerges from a QFT as an effective phenomenon.

Finally, the action in Eq.~(\ref{ym1}) is then mapped into
\begin{equation}
S=\frac{1}{8\pi G}\int\left[\frac{1}{2\Lambda^2}R^\mathfrak{a}_{\phantom{a}\mathfrak{b}}\star R_\mathfrak{a}^{\phantom{a}\mathfrak{b}}+T^\mathfrak{a}\star T_\mathfrak{a}-\frac{\epsilon}{2}\epsilon_\mathfrak{abcd}R^\mathfrak{ab}e^\mathfrak{c}e^\mathfrak{d}+\frac{\Lambda^2}{4}\epsilon_\mathfrak{abcd}e^\mathfrak{a}e^\mathfrak{b}e^\mathfrak{c}e^\mathfrak{d}\right]\;,\label{ym3}
\end{equation}
where $R^\mathfrak{a}_{\phantom{a}\mathfrak{b}}=\mathrm{d}\omega^\mathfrak{a}_{\phantom{a}\mathfrak{b}}+\omega^\mathfrak{a}_{\phantom{a}\mathfrak{c}} \omega^\mathfrak{c}_{\phantom{c}\mathfrak{b}}$ and $T^\mathfrak{a}=\mathrm{d}e^\mathfrak{a}-\omega^\mathfrak{a}_{\phantom{a}\mathfrak{b}}e^\mathfrak{b}$ are the curvature and torsion in $\mathbb{M}^4$, respectively. Moreover, Newton and cosmological constants are obtained from the relations $\gamma^2=\kappa^2/2\pi G$ and $\Lambda^2=\gamma^2/4$.

As a gauge theory in the cotangent bundle \cite{Sobreiro:2010ji,Sobreiro:2012book}, the physical observables have to be gauge invariant operators. In this case, two relatively simple quantities determine the geometry os spacetime, namely, the metric tensor $g=\eta_{ab}e^a\otimes e^b$ and the affine connection $\Gamma^\mu=e^\alpha_a(\delta^a_b\partial_\mu+\omega_{\mu b}^a)e^b_\nu$.

\section{Discussion}

We started with a standard gauge theory in a Euclidean four-dimensional spacetime. The theory is actually renormalizable, at least through all orders in perturbation theory. As a non-Abelian gauge theory, it presents asymptotic freedom and the possibility of dynamical mass generation. Then, a proposition for quantum gravity has been made, as long as it induces an effective geometry that could be interpreted as gravity. The fact that the theory possesses a mass parameter enabled the vierbein to emerge. Moreover, the deformation of the de Sitter algebra at low energies induces a symmetry breaking for the Lorentz group, which finally allows the identification for the fundamental fields with geometric quantities. The resulting effective theory is a geometrodynamical gravity described by action in Eq.~(\ref{ym3}). It is easy to show that the simplest vacuum solution is a de Sitter type spacetime \cite{Sobreiro:2011hb}.

The fact that the gauge group determines the local isometries has a remarkable consequence: For the cases $m\in\{0,1\}$, the reduced group is $SO(4)$ implying that the local isometries are that of an Euclidean space. On the other hand, for $m=2$, it is the Lorentz group $SO(1,3)$ that determines the local isometries. As a consequence of the latter case, space and time are then explicitly distinct from each other. This effect can be interpreted as the rising of the equivalence principle. If unitarity is required for quantum consistency, then a Wick rotation can be attached to the mapping \cite{Sobreiro:2011hb}.

It is also remarkable that Newton's and cosmological constants can be actually computed from the standard quantum field theory techniques, at least at perturbative level. Moreover, they are related quantities through $\Lambda^2=\kappa^2/8\pi G$. Thus, for a small $G$ solution, $\Lambda$ should be big and might compensate for the quantum field theory predictions in order to generate an effective cosmological constant consistent with astrophysical observations.

Nevertheless, one may argue that the presented mechanism violates Weinberg-Witten theorems \cite{Weinberg:1980kq} which forbid: (i) massless charged states with helicity $j>1/2$ that have conserved Lorentz-covariant current and (ii) massless states with helicity $j>1$ that have conserved Lorentz-covariant energy-momentum tensor. However, this is not the case here. First, the theory has a few mass parameters and the theorem holds for massless states only. Second, and more important, there are no spin-2 states in this model. The fields are identified with geometry and not with spin-2 composite fields. Gravity emerges as geometrodynamics and not as a field theory for spin-2 particles.

Let us also compare the present mechanism with the standard model. Strong, weak and electromagnetic interactions are described by gauge theories. At high energies, these theories are very similar (except for the gauge groups). At low energies, however, these theories tend to behave in very different ways. While electrodynamics remains essentially in a perturbative regime, weak interactions suffer spontaneous symmetry breaking through the Higgs mechanism. On the other hand, quark-gluon confinement show up in chromodynamics, and hadronization phenomena take place. Specifically, confinement and the gauge principle state that physical observables must be gauge invariant and colorless. Those states are recognized as hadrons and glueballs. Now, if the present theory can describe gravity, then: (i) at high energies, gravity is a gauge theory very similar to the other fundamental interactions; (ii) at low energies, instead of hadrons and glueballs, the physical observables are identified with geometry, and spacetime itself is affected by this  theory. Thus, geometry appears as the low energy manifestation of gravity, in the same way that hadronization and spontaneous symmetry breaking are the low energy manifestations of chromodynamics and weak interactions.

Finally, for now, we can only say that a standard four-dimensional renormalizable Yang-Mills theory can generate a gravity theory at low energy regime. Obviously, many computations and tests must be performed before we recognize this theory (or some variation) as \emph{the} quantum gravity theory or only an academic exercise.

\section*{Acknowledgements}

R.~F.~S.~acknowledges the organizing committee for the kind invitation to deliver this seminar at the 7ICMMP. Prof.~H.~S.~Smith is acknowledged for reviewing the text. Conselho Nacional de Desenvolvimento Cient\'{i}fico e Tecnol\'{o}gico\footnote{RFS is a level PQ-2 researcher under the program \emph{Produtividade em Pesquisa}, 304924/2009-1.} (CNPq-Brazil) and the Coordena\c{c}\~ao de Aperfei\c{c}oamento de Pessoal de N\'{\i}vel Superior (CAPES) are acknowledged for financial support.


\begin{thebibliography}{99}

\bibitem{Utiyama:1956sy} R.~Utiyama,
{\it Invariant theoretical interpretation of interaction,}
Phys.\ Rev.\  {\bf 101} (1956) 1597. 

\bibitem{Kibble:1961ba}
T.~W.~B.~Kibble, 
{\it Lorentz invariance and the gravitational field,}
J.\ Math.\ Phys.\  {\bf 2} (1961) 212. 
  
\bibitem{Zanelli:2005sa} J.~Zanelli,
{\it Lecture notes on Chern-Simons (super-)gravities,} 2nd ed.~(2008), hep-th/0502193.

\bibitem{'tHooft:1974bx} 
  G.~'t Hooft and M.~J.~G.~Veltman,
  {\it One loop divergencies in the theory of gravitation,}
  Annales Poincare Phys.\ Theor.\ A {\bf 20} (1974) 69.
  
\bibitem{MacDowell:1977jt} 
  S.~W.~MacDowell and F.~Mansouri,
 {\it Unified Geometric Theory of Gravity and Supergravity,}
  Phys.\ Rev.\ Lett.\  {\bf 38} (1977) 739 
  [Erratum-ibid.\  {\bf 38} (1977) 1376].

\bibitem{Stelle:1979aj}
  K.~S.~Stelle and P.~C.~West,
  {\it Spontaneously Broken De Sitter Symmetry And The Gravitational Holonomy Group,}
  Phys.\ Rev.\  {\bf D21} (1980) 1466.
  
\bibitem{Mahato:2004zi}
  P.~Mahato,
{\it De Sitter group and Einstein-Hilbert Lagrangian,}
  Phys.\ Rev.\  {\bf D70} (2004) 124024 [gr-qc/0603100].
  
\bibitem{Tresguerres:2008jf}
  R.~Tresguerres,
  {\it Dynamically broken Anti-de Sitter action for gravity,}
  Int.\ J.\ Geom.\ Meth.\ Mod.\ Phys.\  {\bf 5} (2008) 171-183 [arXiv:0804.1129 [gr-qc]].

\bibitem{Mielke:2010zz}
  E.~W.~Mielke,
  {\it Einsteinian gravity from a spontaneously broken topological BF theory,}
  Phys.\ Lett.\  {\bf B688} (2010) 273-277.
  
\bibitem{Gross:1973id}
  D.~J.~Gross and F.~Wilczek,
 {\it Ultraviolet behavior of non-abelian gauge theories,}
  Phys.\ Rev.\ Lett.\  {\bf 30} (1973) 1343.
  
\bibitem{Politzer:1973fx}
  H.~D.~Politzer,
 {\it Reliable perturbative results for strong interactions?,}
  Phys.\ Rev.\ Lett.\  {\bf 30} (1973) 1346.
  
\bibitem{Dudal:2005na} 
  D.~Dudal, R.~F.~Sobreiro, S.~P.~Sorella and H.~Verschelde,
  {\it The Gribov parameter and the dimension two gluon condensate in Euclidean Yang-Mills theories in the Landau gauge,}
  Phys.\ Rev.\ {\bf D72} (2005) 014016 [hep-th/0502183].
  
\bibitem{Dudal:2011gd} 
  D.~Dudal, S.~P.~Sorella and N.~Vandersickel,
  {\it The dynamical origin of the refinement of the Gribov-Zwanziger theory,}
  Phys.\ Rev.\ {\bf D84}  (2011) 065039 [arXiv:1105.3371 [hep-th]].

\bibitem{Sobreiro:2011hb} 
  R.~F.~Sobreiro, A.~A.~Tomaz and V.~J.~V.~Otoya,
  {\it de Sitter gauge theories and induced gravities,}
  Eur.\ Phys.\ J.\ {\bf C72} (2012) 191 [arXiv:1109.0016 [hep-th]].
  
\bibitem{Inonu:1953sp}
  E.~In\"on\"u and E.~P.~Wigner,
  {\it On the Contraction of groups and their represenations,}
  Proc.\ Nat.\ Acad.\ Sci.\  {\bf 39} (1953) 510-524.
  
\bibitem{Sobreiro:2010ji} 
  R.~F.~Sobreiro and V.~J.~Vasquez Otoya,
  {\it On the topological reduction from the affine to the orthogonal gauge theory of gravity,}
  J.\ Geom.\ Phys.\  {\bf 61} (2011) 137 [arXiv:1003.1102 [hep-th]].
  
\bibitem{Sobreiro:2012book} 
R.~F.~Sobreiro, Ch. 5 {\it Fiber bundles, gauge theories and gravity}
  in  {\it Quantum Gravity}, Ed.~R.~F.~Sobreiro, InTech, Rijeka 2012. 

\bibitem{Weinberg:1980kq} 
  S.~Weinberg and E.~Witten,
  {\it Limits on Massless Particles,}
  Phys.\ Lett.\ {\bf B96} (1980) 59.

\end{thebibliography}
\end{document}